\documentclass{iaus}
\usepackage{graphicx}
\usepackage{amsmath}
\usepackage{natbib}

\title[Equatorial mass loss from Be stars]
{Equatorial mass loss from Be stars}

\author[C. Georgy, S. Ekstr\"om, A. Granada \and G. Meynet]   
{Cyril Georgy$^1$, Sylvia Ekstr\"om$^1$, Anah\'i Granada$^{1,2}$\\ \and Georges Meynet$^1$}

\affiliation{$^1$Geneva University Observatory\\Chemin des Maillettes 51, 1290 Versoix, Switzerland \\[\affilskip]
$^2$Fac. de Cs. Astr. y Geof. Universidad Nacional de La Plata - IALP CCT La Plata,\\UNLP-CONICET, Argentina}

\pubyear{2010}
\volume{272}  
\pagerange{1-2}
\jname{Active OB stars Ð structure, evolution, mass loss, and critical limits}
\editors{C. Neiner, G. Wade, G. Meynet \& G. Peters}
\begin{document}

\maketitle

\begin{abstract}
Be stars are thought to be fast rotating stars surrounded by an equatorial disc. The formation, structure and evolution of the disc are still not well understood. In the frame of single star models, it is expected that the surface of an initially fast rotating star can reach its keplerian velocity (critical velocity). The Geneva stellar evolution code has been recently improved, in order to obtain some estimates of the total mass loss and of the mechanical mass loss rates in the equatorial disc during the whole critical rotation phase. We present here the first results of the computation of a grid of fast rotating B stars evolving towards the Be phase, and discuss the first estimates we obtained.

\end{abstract}

\firstsection 
\section{Introduction}

The Be star phenomenon can be explained in terms of a fast rotating star surrounded by an equatorial disc \citep{Porter2003a}. In this context, the disc formation occurs when the stellar surface reaches (or, at least, becomes close to) the critical rotation \citep[see \textit{e.g.}][]{ekstrom2008b}. When this occurs, the equatorial mass loss is enhanced by the very low effective gravity at the equator, due to the strong centrifugal acceleration.

Recently, the Geneva stellar evolution code was modified in order to account for the equatorial mass loss when the star reaches the critical velocity. If the surface of the star becomes over-critically rotating because of evolutive processes, the extra angular momentum is evacuated. The mass which is removed is estimated assuming that the mass decouples from the star at the surface in the equatorial plane.

In this work, we present the first results of the computation of a grid of B type stars rotating models, in the mass range from $3$ to $15\,\text{M}_\odot$ with three different initial rotation velocities.

\begin{figure}[h!]
\begin{center}
\includegraphics[width=.45\textwidth]{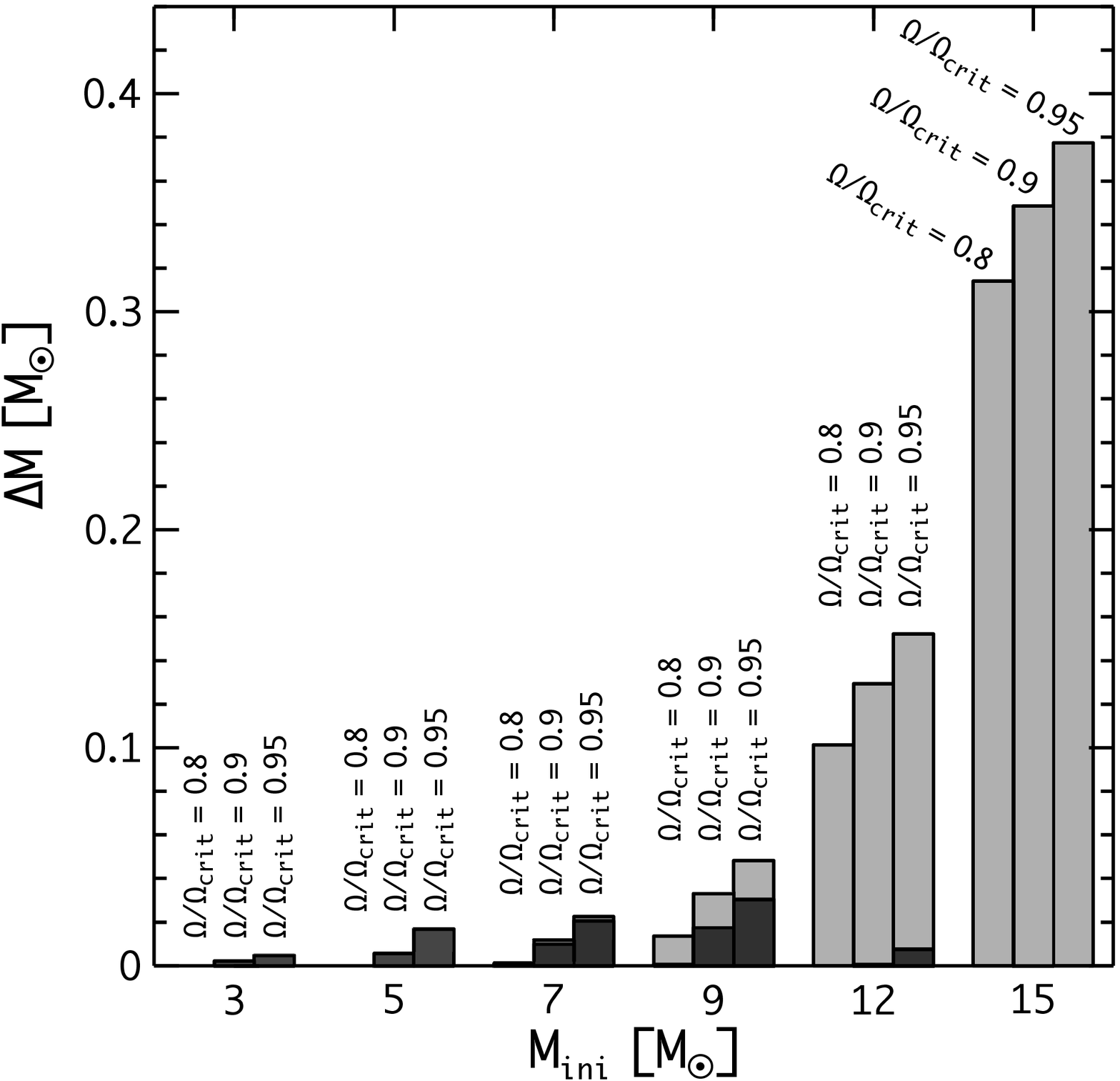}\hfill\includegraphics[width=.45\textwidth]{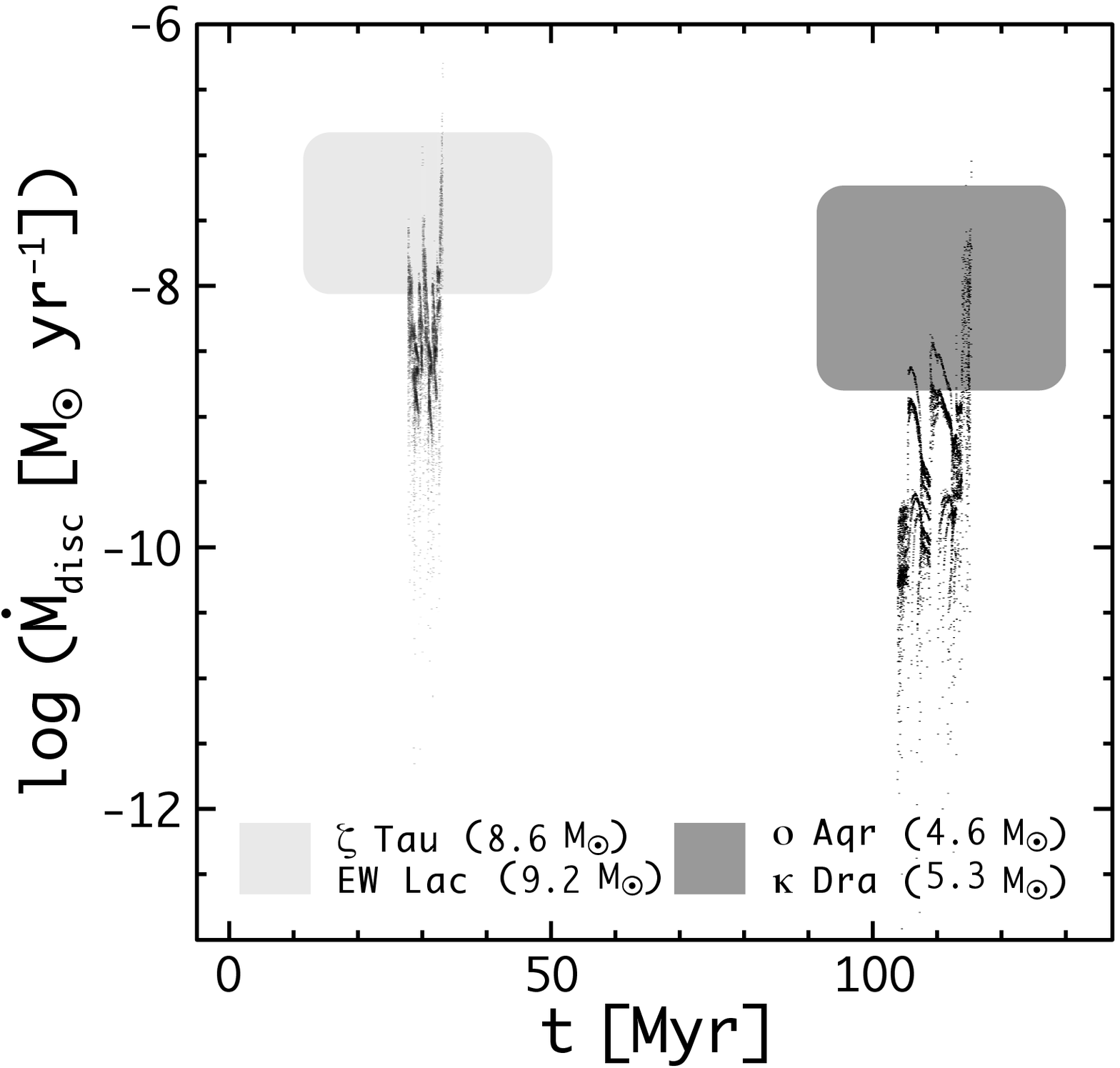}
\caption{\textit{Left panel}: equatorial mechanical mass loss (dark grey) and radiative mass loss (light grey) as a function of the initial mass and initial rotation velocity. \textit{Right panel:} Comparison of our estimated equatorial mass loss rates with observations for the $5\,\text{M}_\odot$ (dark grey points) and $9\,\text{M}_\odot$ (light grey points). The shaded areas correspond to the observed mass loss rates of Be stars with similar masses \citep{Rinehart1999a,Stee2003a,Zorec2005a}}
\label{fig1}
\end{center}
\end{figure}

\begin{table}[h!]
\centering
\caption{Main characteristics of the Be star models. $M_\text{ini}$ is the initial mass of the model; $\Omega/\Omega_\text{crit}$ is the initial rotation parameter; $\Delta M_\text{tot}$ is the total amount of mass lost by the star during the MS; $\Delta M_\text{mec}$ is the amount of mass lost mechanically in the equatorial disc during the critical-rotation phase; $<\dot{M}_\text{mec}>$ is the mean equatorial mass loss rate during the phase of critical rotation; $\tau_\text{cr}/\tau_\text{MS}$ is the fraction of the MS spent at the critical velocity; $X_\text{C,cr}$ is the fraction of hydrogen in the burning core when the star reaches the critical velocity for the first time; $<v_\text{eq}>$ is the mean equatorial velocity during the MS.}
\scriptsize{
\begin{tabular}{cc|cccccc}
\hline
    $M_\text{ini}\,[\text{M}_\odot]$ & $\Omega/\Omega_\text{crit}$ & $\Delta M_\text{tot}\,[\text{M}_\odot]$ & $\Delta M_\text{mec}\,[\text{M}_\odot]$ & $<\dot{M}_\text{mec}>\,[\text{M}_\odot\,\text{yr}^{-1}]$ & $\tau_\text{cr}/\tau_\text{MS}$ & $X_\text{C,cr}$ & $<v_\text{eq}>\,[\text{km}\,\text{s}^{-1}]$\\
\hline
    $3$ & $0.8$ & $0.0$ & $0.0$ & $0.0$ & $0.0$ & $-$ & $219.7$\\
    & $0.9$ & $2.36\cdot 10^{-3}$ & $2.36\cdot 10^{-3}\,(100\%)$ & $5.6\cdot 10^{-11}$ & $0.10$ & $0.16$ & $265.8$\\
    & $0.95$ & $5.49\cdot 10^{-3}$ & $5.49\cdot 10^{-3}\,(100\%)$ & $4.7\cdot 10^{-11}$ & $0.27$ & $0.33$ & $290.8$\\
    $5$ & $0.8$ & $0.0$ & $0.0$ & $0.0$ & $0.0$ & $-$ & $248.3$\\
    & $0.9$ & $6.34\cdot 10^{-3}$ & $6.34\cdot 10^{-3}\,(100\%)$ & $5.5\cdot 10^{-10}$ & $0.10$ & $0.16$ & $300.5$\\
    & $0.95$ & $1.91\cdot 10^{-2}$ & $1.91\cdot 10^{-2}\,(100\%)$ & $4.0\cdot 10^{-10}$ & $0.38$ & $0.40$ & $349.4$\\
    $7$ & $0.8$ & $1.55\cdot 10^{-3}$ & $4.93\cdot 10^{-4}\,(3\%)$ & $1.1\cdot 10^{-9}$ & $0.001$ & $0.0$ & $268.3$\\
    & $0.9$ & $1.33\cdot 10^{-2}$ & $1.13\cdot 10^{-2}\,(85\%)$ & $1.5\cdot 10^{-9}$ & $0.14$ & $0.20$ & $323.6$\\
    & $0.95$ & $2.57\cdot 10^{-2}$ & $2.34\cdot 10^{-2}\,(91\%)$ & $1.6\cdot 10^{-9}$ & $0.27$ & $0.31$ & $356.0$\\
    $9$ & $0.8$ & $1.53\cdot 10^{-3}$ & $7.82\cdot 10^{-4}\,(5\%)$ & $6.2\cdot 10^{-9}$ & $0.004$ & $0.01$ & $286.5$\\
    & $0.9$ & $3.75\cdot 10^{-2}$ & $1.98\cdot 10^{-2}\,(53\%)$ & $3.5\cdot 10^{-9}$ & $0.17$ & $0.22$ & $347.7$\\
    & $0.95$ & $5.48\cdot 10^{-2}$ & $3.45\cdot 10^{-2}\,(63\%)$ & $3.3\cdot 10^{-9}$ & $0.31$ & $0.33$ & $377.0$\\
    $12$ & $0.8$ & $1.15\cdot 10^{-1}$ & $0.0\,(0\%)$ & $0.0$ & $0.0$ & $-$ & $300.3$\\
    & $0.9$ & $1.47\cdot 10^{-1}$ & $9.08\cdot 10^{-4}\,(1\%)$ & $3.2\cdot 10^{-10}$ & $0.14$ & $0.18$ & $363.2$\\
    & $0.95$ & $1.73\cdot 10^{-1}$ & $8.56\cdot 10^{-3}\,(5\%)$ & $1.5\cdot 10^{-9}$ & $0.27$ & $0.29$ & $394.3$\\
    $15$ & $0.8$ & $3.57\cdot 10^{-1}$ & $0.0\,(0\%)$ & $0.0$ & $0.0$ & $-$ & $302.5$\\
    & $0.9$ & $3.96\cdot 10^{-1}$ & $0.0\,(0\%)$ & $0.0$ & $0.0$ & $-$ & $353.6$\\
    & $0.95$ & $4.29\cdot 10^{-1}$ & $0.0\,(0\%)$ & $0.0$ & $0.0$ & $-$ & $384.4$\\
\hline
\end{tabular}}
\label{tab1}
\end{table}

\section{Preliminary results}

In Fig.~\ref{fig1} (left panel), we show the relative contribution of the equatorial mechanical mass loss during the critical rotation phase compared to the radiative mass loss. For the less massive stars, the mechanical mass loss is the dominant process in these fast rotating models. The radiative mass loss becomes ever larger for higher mass stars, preventing progressively the star to reach the critical velocity. In the right panel, we compare our $5$ and $9\,\text{M}_\odot$ models with observed equatorial mass loss rates from Be stars. We see that our results are marginally compatible with the observations. Note however that the observed values are instantaneous values, whereas the computed ones are averaged values, due to the quite long time step needed for the stellar evolution computations. The main properties of our models of Be stars are summarised in Table~\ref{tab1}.

\bibliographystyle{cup}
\bibliography{s1-03_georgy_biblio}

\end{document}